\documentclass[twocolumns]{aastex631}

\usepackage{graphicx}	% Including figure files
\def\ltsima{$\; \buildrel < \over \sim \;$}
\def\lsim{\lower.5ex\hbox{\ltsima}}

\shorttitle{A search for soft emission lines in GRB 221009A}
\shortauthors{Campana et al.}

\begin{document}

\title{A search for soft X--ray emission lines in the afterglow spectrum of GRB 221009A}

\author[0000-0001-6278-1576]{Sergio~Campana}
\affiliation{INAF--Osservatorio Astronomico di Brera, Via Bianchi 46, I-23807, Merate (LC), Itay}

\author{Valentina Braito}
\affiliation{INAF--Osservatorio Astronomico di Brera, Via Bianchi 46, I-23807, Merate (LC), Itay}
\affiliation{Department of Physics, Institute for Astrophysics and Computational Sciences, The Catholic University of America, Washington, DC 20064, USA}
\affiliation{Dipartimento di Fisica, Universit\`a di Trento, Via Sommarive 14, I-38123, Trento, Italy}

\author{Davide Lazzati}
\affiliation{Department of Physics, Oregon State University, 301 Weniger Hall, Corvallis, OR 97331, USA}

\author{Andrea Tiengo}
\affiliation{Scuola Universitaria Superiore IUSS Pavia, Palazzo del Broletto, Piazza della Vittoria 15, I-27100 Pavia, Italy}
\affiliation{INAF-Istituto di Astrofisica Spaziale e Fisica Cosmica di Milano, Via A. Corti 12, I-20133 Milano, Italy}

\begin{abstract} 
GRB 221009A was the Brightest gamma-ray burst Of All Time (BOAT), 
surpassing in prompt brightness all GRBs discovered in $\sim 50$ yr and in afterglow brightness in $\sim 20$ yr. We observed the BOAT 
with {\it XMM-Newton} 2.3 d after the prompt. The X--ray afterglow was still very bright and we collected the largest number of photons with the Reflection Grating Spectrometers (RGS) on a GRB. We searched the RGS data for narrow emission or absorption features. We did not detect any bright line feature. A candidate narrow feature is identified at a (rest-frame) energy of $1.455^{+0.006}_{-0.014}$ keV, consistent with an Mg XII $K_\alpha$ emission line, slightly redshifted (0.012) with respect to the host galaxy. 
We assessed a marginal statistical significance of $3.0\,\sigma$ for this faint feature based on conservative Monte Carlo simulations, 
which requires caution for any physical interpretation.
If this line feature would be for real, we propose that it might originate
from the reflection in the innermost regions of the infalling funnel from low-level late-time activity emission of the central engine.
\end{abstract} 


%
%----------------------------------------------------------------

\section{Introduction} \label{sec:intro}

Emission and/or absorption features have been lengthily searched for in the afterglow spectra of Gamma-Ray Bursts (GRBs). If revealed, these lines are a powerful diagnostic of the burst ambient, geometry, and physical conditions which made possible the line emission \citep{Lazzati1999, Lazzati2002, Ballantyne2001, Ghisellini2002, Kallman2003}.
The first putative detections came just after the discovery of the GRB afterglows, with {\it BeppoSAX} (GRB 970508, \citealt{Piro1999}; GRB 000214 \citealt{Antonelli2000}) and {\it ASCA} (GRB 970828, \citealt{Yoshida1999}).
These were all lines consistent with Iron K$\alpha$ emission. 
{\it XMM-Newton} and {\it Chandra} provided detections, too.
An iron line was identified in the {\it Chandra} data (GRB 991216, \citealt{Piro2000}). Reeves et al. (\citeyear{Reeves2002}) detected several emission lines from mid-Z elements (Mg XI, Si XIV, S XVI, Ar XVIII, and Ca XX) in the {\it XMM-Newton}-pn spectrum of GRB 011211. Similar lines were detected in the {\it XMM-Newton} spectra of GRB 001025 \citep{Watson2002} and GRB 030227 \citep{Watson2003}. The presence of a Sulfur line was suggested to be present in the {\it Chandra} HETGS spectrum of the afterglow of GRB 020813 \citep{Butler2003}. 

The statistical significance of these line features has been discussed at length. Iron lines were quoted at a single-trial significance level of $\sim 3\,\sigma$, assuming the line energy. Only the line in the GRB 991216 {\it Chandra} spectrum, was detected at the quoted significance of $4.7\,\sigma$ \citep{Piro2000}. Most of these claims were based on using the F-test. Prompted by the use of the F-test to assess the significance of a line, Protassov et al. (\citeyear{protassov02}) convincingly argued that the F-test cannot be applied to evaluate the statistical significance of an added emission line, being the F-test a global test and the line addition a local improvement. 
To estimate the significance of the lines in GRB 011211, Reeves et al. (\citeyear{Reeves2002}) ran Monte Carlo simulations, estimating a significance of $3.6\,\sigma$ \citep{Reeves2002, Reeves2003}. Borozdin \& Trudolyubov (\citeyear{Borozdin2003}) criticised the background subtraction and \cite{Rutledge2003} the statistical analysis method.

\cite{Sako2005} carried out the most complete and uniform analysis of 21 GRBs for which emission lines were reported, concluding that no credible X--ray line feature has been detected in any of the GRB afterglows. After this work, no further detections of emission-line features in the afterglow spectra of GRBs were reported. Negative results were reported in the first years of {\it Swift} activity \citep{Hurkett2008}. \cite{Campana2016} selected GRB afterglow spectra collected with the Reflection Grating Spectrometers (RGSs) on board {\it XMM-Newton} with a fluence larger than $10^{-7}$ erg cm$^{-2}$ within the {\it XMM-Newton} observation. This criterion allowed us to collect at least 3,000 net counts in the RGS spectra of the GRB afterglows. Seven GRBs were selected. A few line features were revealed and their significance was assessed through detailed Monte Carlo simulations. The statistical significance resulted in $2-3\,\sigma$, and most of the line energies were compatible with Galactic Oxygen K$\alpha$ or Oxygen edge features. These observations taught us that we need the highest possible number of photons in the RGS, achievable by triggering on the brightest GRBs as soon as possible.

Here we approach the Brightest Of All Time (BOAT) GRB 221009A \citep{Burns2023}. This long GRB was first reported as a possible Galactic transient by {\it Swift} \citep{Dichiara2022, Williams2023}. Soon after it was realised that {\it Swift} triggered on the bright afterglow $\sim 55$ min after the GRB prompt emission, observed by {\it Fermi}/GBM \citep{Veres2022, Burns2023}, as well as any other satellite with at least an anti-coincidence shield. We obtained an {\it XMM-Newton} observation through the Director Discretionary Time, with the specific aim of searching for line features in the RGS and for timing features in the EPIC data. The observation started 2.3 d after the GRB, in the first available observing slot.

In Section 2 we describe the data analysis and in Section 3 we describe the Monte Carlo simulations carried out to assess the line significance. In Section 4 we discuss the possible origin of the line. 
Section 5 is dedicated to the Conclusions.

\section{Data Analysis}

\subsection{Data extraction}

The {\it XMM-Newton} observation started on 2022-10-11 UT21:10:14 lasting 45.7 ks (RGS). The two RGS were in standard spectroscopic mode, and the three EPIC cameras were all in timing mode with the thick filter, to minimise the pile-up of the bright afterglow. No background flares were recognised. Data were reduced with the {\it XMM-Newton} Science Analysis Software (SAS) version xmmsas\_20201028\_0905 (SAS 19.0.0) and the latest calibration files. Data were locally reprocessed with the tasks {\tt rgsproc} and {\tt epproc}. RGS data were extracted from a standard region. We visually inspect that the dust scattering rings \citep{Tiengo2023, Williams2023, Vasilopoulos2023} were contaminating the extraction region. The large intrinsic absorption prevented us from attaining a sufficiently large number of photons below 1 keV. The spectral analysis has been carried out in the 1--2 keV energy range. In the combined RGS(1+2) 1--2 keV spectrum there are 12,234 counts. The background comprises 16\% of the total counts. RGS data were rebinned by 16 channels. This is equivalent to rebin the spectra with a constant energy bin of about $\Delta \lambda =0.16$\AA, where 0.1 \AA\ is the spectral resolution of the RGS gratings.

The pn instrument data were taken in timing mode with the thick filter. Given the higher quality of the pn spectrum and the better calibration of the pn with respect to the MOS in timing mode, we did not consider further the MOS data. 
We filtered the pn data by selecting grades 0--4, {\tt FLAG==0} and {\tt \#XMMEA\_EP} options. The pn events were extracted from a four-column wide strip centred on the source (column 38). Background events were extracted in a twice larger region centred on column 7. As can be appreciated by a nearly simultaneous {\it Swift} observation, the presence of the dust scattering rings is apparent at this time \citep{Williams2023}. Timing data do not allow us to isolate the rings along the extraction column, lacking an image. For this specific reason and not to alter the line search in the 1--2 keV energy range, we selected data in the 2--9 keV energy band only, where the contribution from the dust scattering rings is negligible. This will prevent us from confirming RGS lines with the higher signal-to-noise pn spectrum, but, unfortunately, the ring contribution within the pn extraction region cannot be disentangled from the source. We collected 253,814 photons, with the background comprising a mere $3\%$. The count rate is $\sim 5$ cts s$^{-1}$, with no pile-up. Data were rebinned to 16 channels, too, to have a good handle of the spectral shape.

%%%%%%%%%%%%%%%%%%%%%%%%%%%%%%%%%%%%%%%%%%%%%%%%%%%%%
\begin{figure}
\centering
%\plotone{line.jpg}
\includegraphics[width=0.5\textwidth]{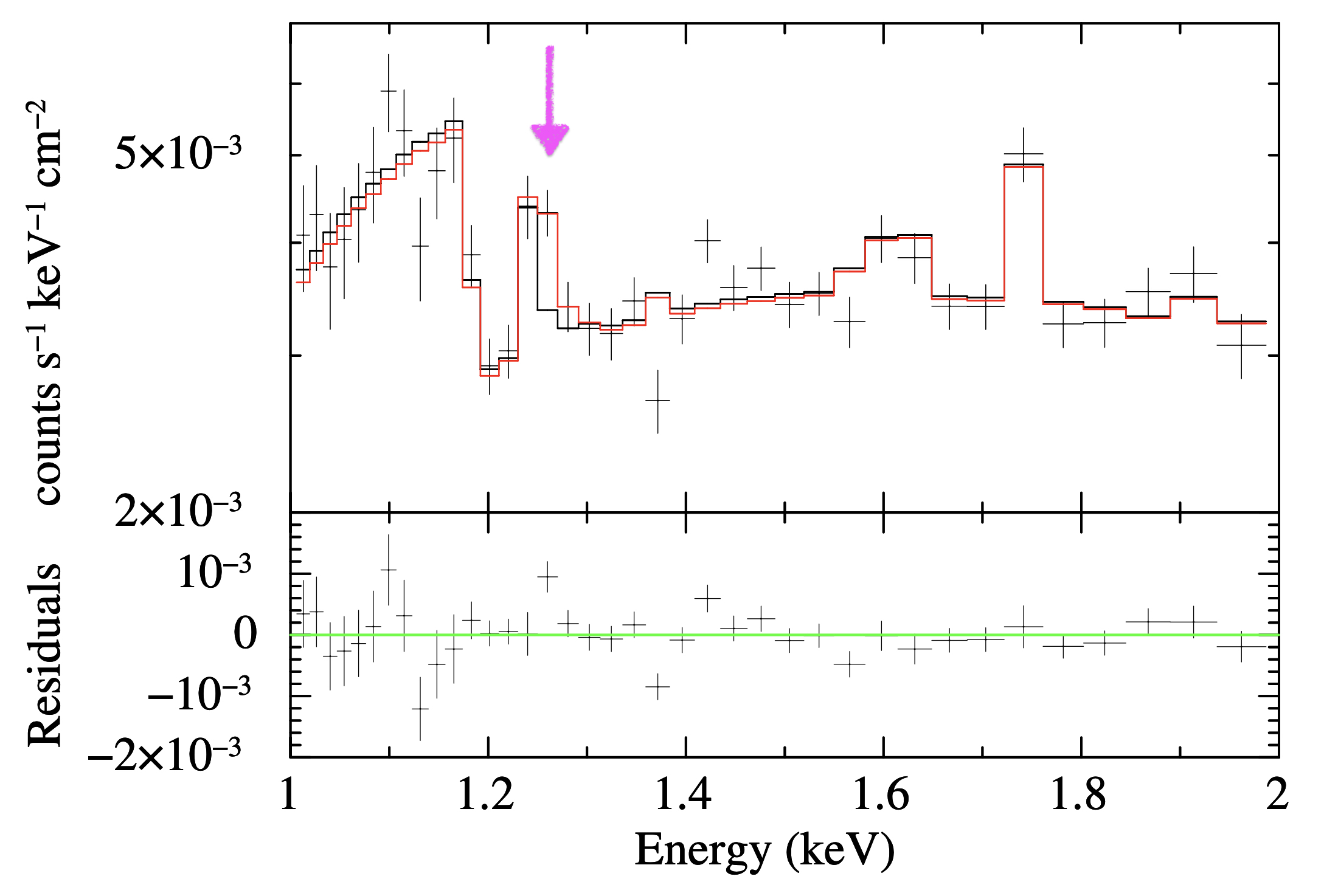}
\caption{RGS spectrum of GRB 221009A. The data have been normalised by the RGS effective area (setplot area in {\tt XSPEC}). The absorbed power-law best-fit model is represented with a continuous black line. The best-fit model including the Mg emission line is over-plotted in red (to show the minimal change of the continuum). A pink vertical arrow indicates the Mg line energy. Residuals for the power law model only are shown in the lower panel. 
\label{fig:line}}
\end{figure}
%%%%%%%%%%%%%%%%%%%%%%%%%%%%%%%%%%%%%%%%%%%%%%%%%%%%%

\subsection{Spectral fitting}

In the early stages of the afterglow, GRB might show flares that considerably alter their X--ray spectra or soft energy components. After $\sim 1.5$ d (rest-frame) the event, however, the afterglow spectra are relatively simple, usually accounted for by an absorbed power-law model. The absorption is made of a Galactic component, an intrinsic component within the host galaxy, and a contribution along the line of sight \citep{Campana2015, Behar2011}.
For GRB 221009A the very close distance ($z=0.151$, \citealt{Malesani2023}), rules out any possible contribution from material along the line of sight.
For such a small redshift, the Galactic and host galaxy $N_H$ contribution is almost interchangeable. Taking advantage of three further {\it XMM-Newton} observations collected in the following days totalling $\sim 78,000$ counts with the pn and $\sim 58,000$ with the two MOSs, we were able to disentangle the Galactic and host galaxy contribution (Campana et al. in preparation). This allowed us to fix the value of the Galactic column density to $9\times 10^{21}$ cm$^{-2}$. This value is higher than the value derived by the "Willingale" tool \citep{Willingale2013} of $5.3\times 10^{21}$ cm$^{-2}$, but it is in line with the map of total hydrogen column density in the sky area around GRB 221009A derived by Planck \citep{Planck2014} and with the value derived from the dust scattering ring analysis (see Fig. 4 in \cite{Tiengo2023}).

\begin{table}
\begin{center}
\caption{Results of the GRB 221009A afterglow spectral fit.}
\begin{tabular}{cc}
\hline 
Parameter & Value \\
\hline
$N_H({\rm Gal}$ ($10^{21}$ cm$^{-2}$) & 0.9 (fixed)\\
$N_H(z)$ ($10^{21}$ cm$^{-2}$) & $4.8\pm{0.8}$ \\
Photon index ($\Gamma$) & $1.72\pm0.01$\\
Normalisation & $(1.48\pm0.07)\times 10^{-2}$\\
Constant$^*$ & $0.99\pm0.03$\\
Flux$^+$ & $5.3^{+0.2}_{-0.3}$\\
\hline 
\end{tabular}
\begin{flushleft}
\small 
\noindent Errors are $90\%$ confidence level for one parameter of interest (i.e., $\Delta$C=2.71).\\
\noindent $^*$ The constant refers to the pn instrument, kept 1 the value for RGS1+2.\\
\noindent Unabsorbed flux in the 0.3--10 keV energy range in units of $10^{-11}$ erg cm$^{-2}$ s$^{-1}$.
\end{flushleft}
\end{center}
\end{table}

\begin{figure}
\begin{center}
\includegraphics[width=0.52\textwidth]{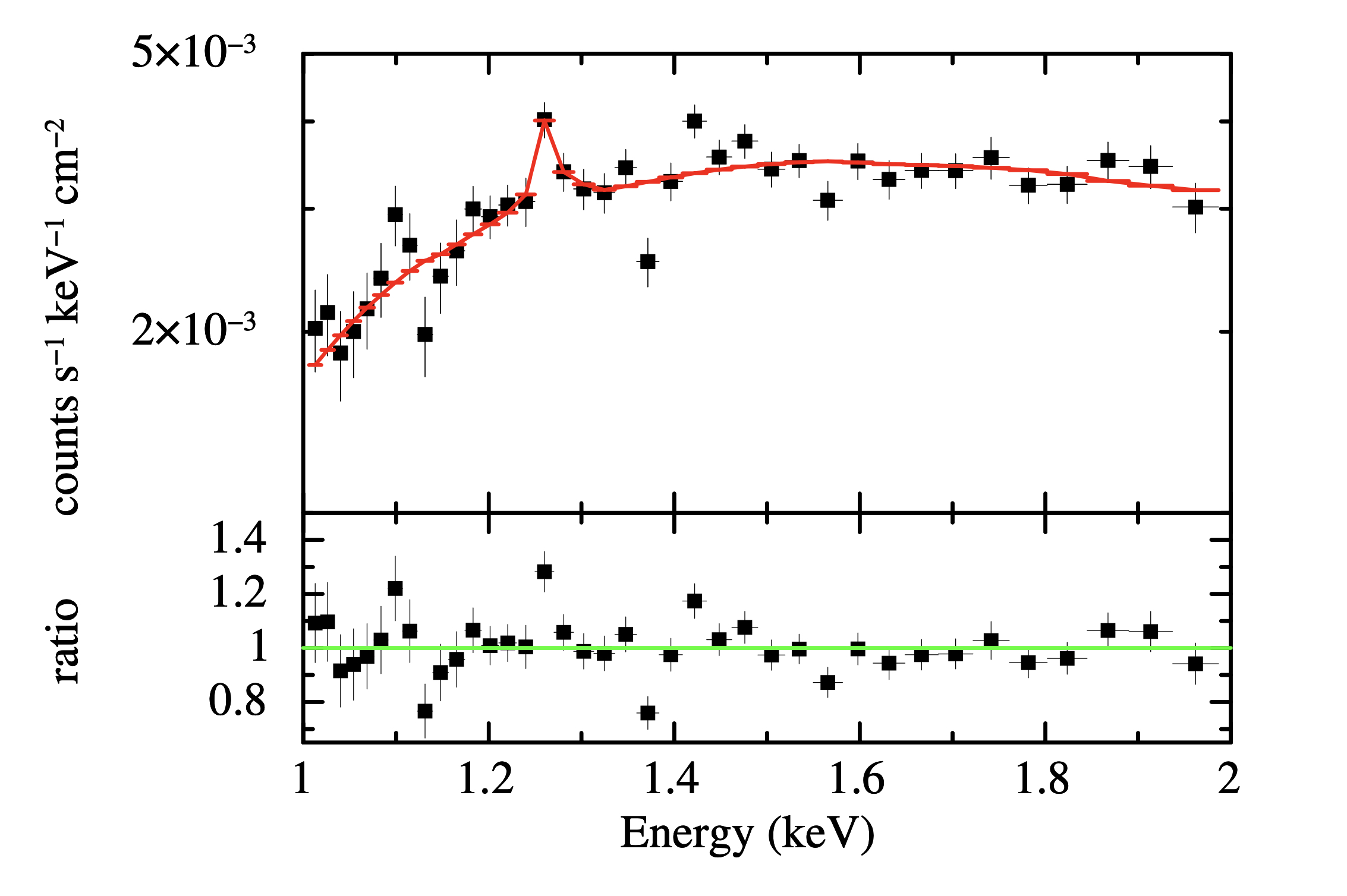}
\includegraphics[width=0.51\textwidth]{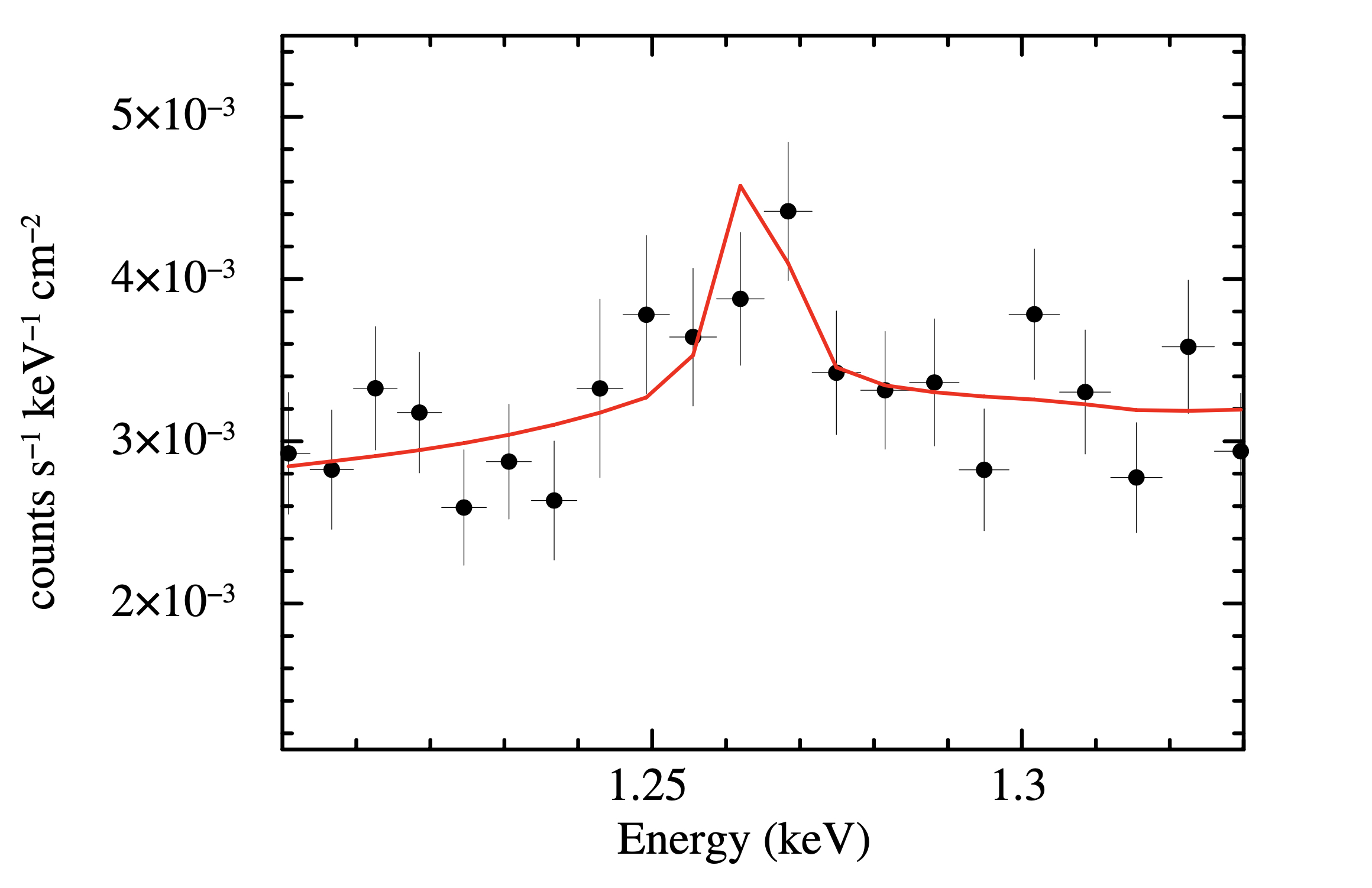}
\includegraphics[width=0.49\textwidth]{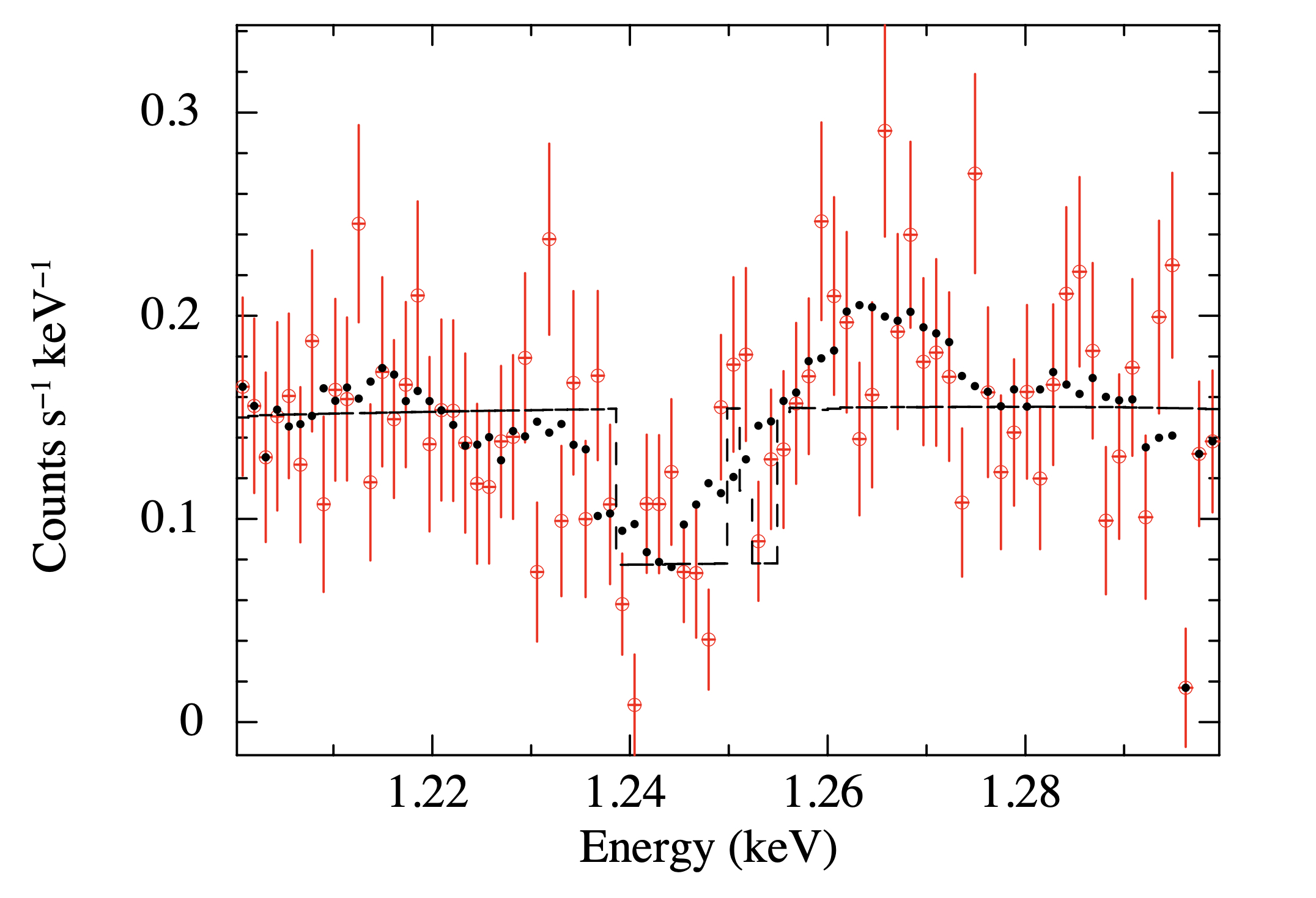}
\caption{Upper panel: unfolded RGS spectrum of GRB 221009A. The best-fit absorbed power-law best-fit model including the Mg emission line is depicted in red. The ratio plot in the lower panel shows the ratio without the line.
Medium panel: zoom of the line energy range with an RGS spectrum rebinned by only a factor of 3 in channels. The best-fit model is in red.
Lower panel: original unbinned data (open red points) in the 1.2--1.3 keV together with smoothed (but not binned) data (filled black points) with a Gaussian kernel of 7 eV, equivalent to the RGS FWHM at these energies. The dashed line shows the spectral model with no line. The large dip between 1.24-1.25 keV and the small dip just above 1.25 keV are due to the instrument response.
} 
\label{fig:ufs}
\end{center}
\end{figure}

We fitted the joined RGS1+2 (1--2 keV) and pn (2--9 keV) spectra with an absorbed power-law model, including an inter-calibration constant with the {\tt XSPEC} package v. 12.13.0c \citep{Arnaud1996}. This spectral model represents our baseline model. We also tested for the presence of a high-energy cut-off with negative results. We adopted C-statistics. We obtain a C-stat value of 152.55 with 123 degrees of freedom (dof). Best fit values are reported in Table 1.
In Fig. \ref{fig:line} we report the residuals to this baseline continuum model; it is apparent the presence of possible line-like residuals, both in emission and in absorption. Therefore we included a narrow line, either in emission or absorption. The line width has been kept fixed below the instrumental resolution (1 eV width). The line energy has been searched for stepping the energy in steps of 1 eV across the entire 1--2 keV energy range. Having found the best local minimum, we optimised our best-fit search in this interval. The best-fit rest-frame line energy is $1.455^{+0.006}_{-0.014}$ keV (all errors are $90\%$ confidence level for one parameter of interest; i.e., $\Delta$C=2.71, otherwise stated). The line equivalent width is 9.7 eV. 
The line is unresolved at the RGS spectral resolution. Leaving free the line width remains constrained to 0 eV, with an upper limit of 15.6 eV.
At the line energy, the continuum is 4.4 times more intense than the line. The ratio of the 0.3--10 keV line flux to the continuum flux is $0.2\%$ (see Fig. \ref{fig:ufs}). 
The line luminosity is estimated to be $\sim(12\pm6)\times 10^{42}$~erg s$^{-1}$ (90\% confidence level).
The addition of the line improves the fit to a C-stat of 135.02 (121 dof), with a $\Delta C=17.53$. Column density and power law parameters are basically unaffected (see Fig. \ref{fig:line}). We searched for other possible line features. The residuals show a weak absorption feature at 1.37 keV (rest-frame). However, the addition of a second line improves the fit by only $\Delta C=12.5$. A further absorption line is at 1.58 keV (rest-frame) producing still a lower $\Delta C=10.7$.

To account for the small difference in the number of degrees of freedom (123 vs. 121), we also consider the Bayesian Information Criterion (BIC) as a likelihood measure penalised for model complexity for comparing the two models (i.e. without and with line). We obtained $\Delta({\rm BIC})=9.9$. A widely accepted threshold for accepting the model is $\Delta({\rm BIC})>8-10$ \citep{Kass1995}.

The closest line to an energy of 1.455 keV is Mg XII (1s-2p, L$\alpha$) at 1.473 keV. 
The line is therefore slightly redshifted with respect to the laboratory value by $z_{\rm shift}=0.012^{+0.010}_{-0.004}$ (i.e. at a redshift of 0.163).
Other lines possibly present in the 1--2 keV energy range of the RGS are Mg XI (1.352 keV), Si XIII (1.864 keV), and Si XIV L$\alpha$ (2.006 keV). For all of them, we fix the line energy at the host galaxy's redshift value or at the Mg XII redshift value, and the line width at 1 eV, and we obtain normalisation consistent with zero in all cases. Note that the most prominent Oxygen and Neon lines do not fall within the searched energy range: within the 1--2 keV energy range we might expect only Magnesium and Silicon lines. We also looked for an iron line in the pn data in the rest frame interval 6.4-6.93 keV. We obtained a $3\,\sigma$ upper limit of 4 eV on the line equivalent width. 
Finally, we tried to add an {\tt APEC} model, with minimal improvement to the fit, even allowing for a free abundance and a redshift interval around the host galaxy value.

Even if the EPIC pn spectrum is contaminated by the dust rings, the pn is sensitive to the 1--2 keV energy band and it has much larger statistics than the RGS. If we include this energy range of the pn data into the fit, the line energy decreases to $1.447^{+0.012}_{-0.008}$ keV, i.e. slightly shifted toward smaller energies. As expected, the line EW is smaller (4.4 eV) and the 
$\Delta C=13.0$, and, more importantly, the $90\%$ confidence level for the line normalisation is still larger than zero at 
$\sim 3.6\,\sigma$.

A second {\it XMM-Newton} observation was carried out 4.4 d (rest-frame) after the burst \citep{Tiengo2023}. In that 60 ks observation, the Mg XII line is not detected, with a $3\,\sigma$ upper limit on the luminosity of $\lsim 7\times 10^{42}$ erg s$^{-1}$ (at 720 Mpc). 

\section{Assessing the statistical significance}

Following the work of \cite{protassov02}, we assess the line significance through detailed Monte Carlo simulations.
We first binned the RGS spectrum to 100 counts per energy bin ($84\%$ due to the source) and verified that the statistics of the $\Delta\chi$ residuals in the 1--2 keV energy band follow a Gaussian distribution.

The null hypothesis model is the best-fit model for the RGS and pn spectra together with the associated background spectra, with no emission or absorption lines over the entire RGS energy range of 1--2 keV. We simulated 10,000 combined spectra with the {\tt fakeit} command in {\tt XSPEC} and grouped the data as for the original observations. The intrinsic column density and the power law photons index were allowed to vary within the combined $3\,\sigma$ uncertainties ($\Delta$C=10.96).
We fit the simulated spectra with the null hypothesis model. Based on this fit, we then simulated again the spectra using this best-fit model as input to assure us that we sample the ``true'' model from which our observation was derived.
This two-step simulation process, besides lengthening the computation time, is particularly effective in driving the fit out of the original best-fit model, exploring a very large space of parameters. This is very conservative too: limiting to the first simulation step the line significance would have increased severely. 

On the second simulated set of spectra, we searched for an emission or absorption Gaussian line across the 1--2 keV energy range. As for the line to be tested, the Gaussian line was assumed to be unresolved (1 eV). We stepped the energy centroid in increments of 20 eV, refitting the spectra at each step. After this process, we recorded the best (lowest) $C$ statistic value for each of the simulated sets of spectra and constructed a distribution of the improvements ($\Delta C$) with respect to the null hypothesis model for each simulation.
The distribution of line energies covers the entire energy range with no particular accumulation points. 
From the distribution of 10,000 simulated spectra, we obtained 28 instances with an improvement in statistics for the inclusion of a line better than the observed value of 17.53. This equates to a Gaussian statistical significance of $3.0\,\sigma$. 
Admittedly, the line significance is not overwhelming, which is too low to derive a firm conclusion on the presence of this line.

%%%%%%%%%%%%%%%%%%%%%%%%%%%%%%%%%%%%%%%%%%%%%%%%%%%%%
%%%%%%%%%%%%%%%%%%%%%%%%%%%%%%%%%%%%%%%%%%%%%%%%%%%%%

%%%%%%%%%%%%%%%%%%%%%%%%%%%%%%%%%%%%%%%%%%%%%%%%%%%%%
\begin{figure}[!ht]
	\centering
\includegraphics[width=\columnwidth]{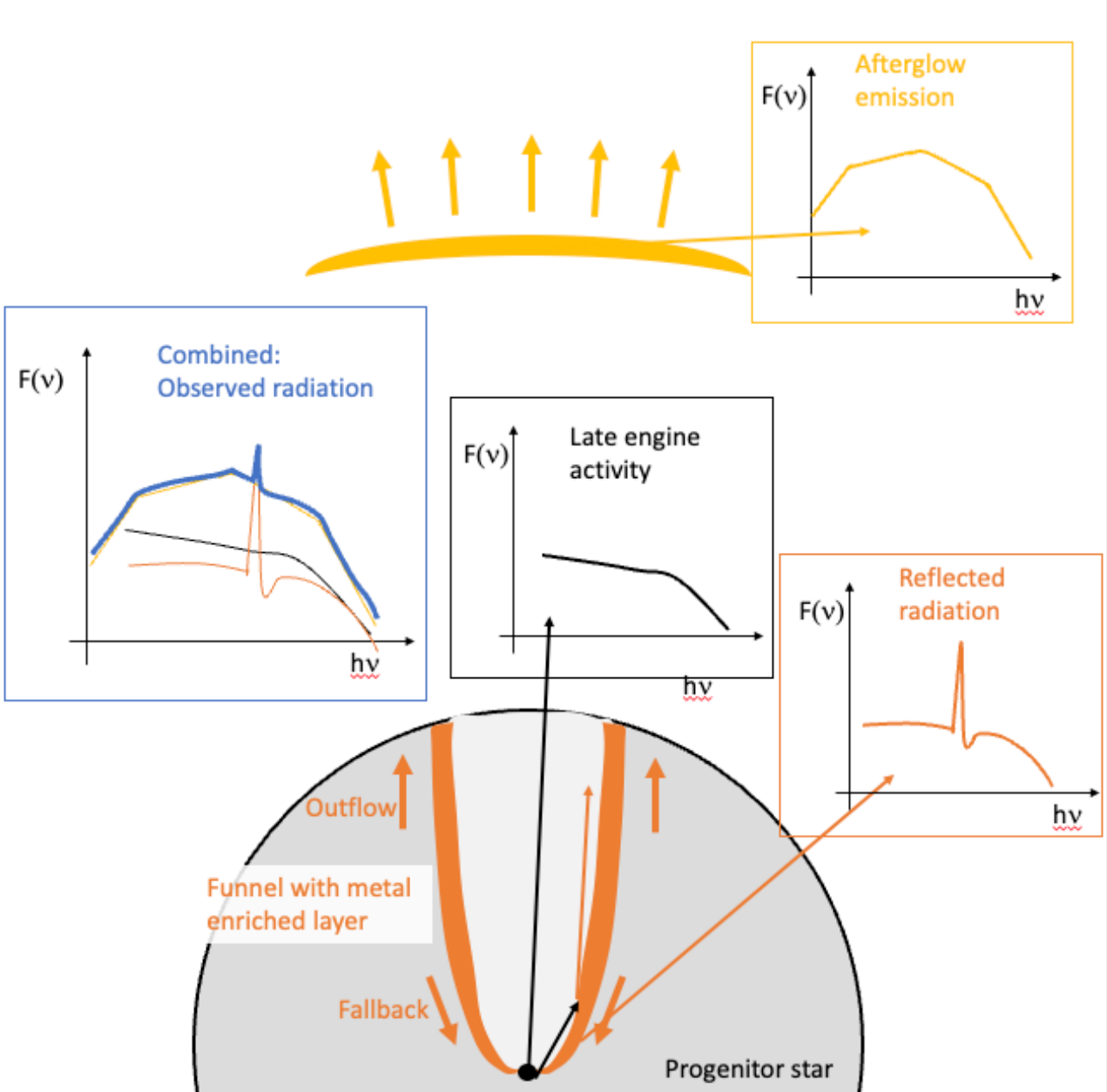}
\caption{Cartoon of the scenario for the production of the redshifted MgXII emission line.} 
\label{fig:cartoon}
\end{figure}
%%%%%%%%%%%%%%%%%%%%%%%%%%%%%%%%%%%%%%%%%%%%%%%%%%%%%

\section{Explaining the putative Mg emission line}

If the Mg XII line would be for real, it cannot be produced by the burst fireball itself given that no significant blueshift is detected, while the fireball is expected to still be relativistic a few days after the burst explosion. Most models for soft X--ray emission lines in GRB afterglows rely on reflecting X--ray continuum off a dense, metal-rich slab \citep{Rees2000, Lazzati2002}. The funnel of a recently exploded collapsar \citep{Woosley2006} provides a natural location for the reflection (see Fig.~\ref{fig:cartoon}). It is compact and dense, characterised by non-relativistic motion, and possibly metal-enriched if the jet propagation dredges core material from the progenitor star \citep{Mazzali2005}. The illuminating continuum, however, cannot be the burst afterglow itself, which is produced approximately a parsec away from the exploding star. A possible source of the local illuminating continuum is a late activity from the burst engine. This is not unlikely
since a similar scenario is invoked to explain the so-called X--ray flares which are seen in a significant fraction of burst afterglows \citep{Lazzati2008, Margutti2011} or the X--ray plateau afterglow phase (e.g., \citealt{Lyons2010, Metzger2011}).

Back of the envelope calculations show that if the central engine remains active after the initial high-luminosity phase, it can produce a featureless continuum (the black SED in Fig.~\ref{fig:cartoon}) that reflects off the funnel walls (the orange SED in Fig.~\ref{fig:cartoon}). Some of the star material should not attain escape velocity and fall back onto the collapsing stellar core, explaining the observed redshift in the line frequency. 

\section{Discussion and conclusions}

We carried out a thorough search for emission or absorption narrow line features in the RGS (1--2 keV) spectrum of the BOAT, GRB 221009A. We did not detect any bright line feature. We revealed only a possible emission line in the afterglow spectrum. The line significance in this  GRB has been assessed through detailed Monte Carlo simulations, resulting in $3.0\,\sigma$. The line energy is (almost) consistent with the Mg XII transition, allowing for a small additional redshift of 0.012. 
A second {\it XMM-Newton} observation 4.4 d after the GRB explosion failed to detect the line. This might indicate that the line is not real, but since the late engine activity will likely decrease across the 2.0-4.4 d interval, also this late non-detection is inconclusive.

We also present a scenario for the formation of emission features with the properties of this Mg line. It invokes extended emission from the central engine up to a few days after the burst, producing the X--ray continuum that illuminates the walls of the funnel produced by the GRB jet inside the progenitor star. The funnel walls, metal-enriched with material from the progenitor core, reflect the continuum, producing soft X--ray emission lines. 

To firmly detect and study these lines, we will need facilities with large effective area telescopes coupled with grating and/or calorimeter X--ray detectors, such as {\it Athena}.

\begin{acknowledgments}
SC acknowledges funding from the Italian Space Agency, contract ASI/INAF n. I/004/11/4. DL acknowledges support from NSF grant AST-1907955. We thank Norbert Schartel for awarding this DDT observation and the {\it XMM-Newton} staff for its swift execution. We thank the anonymous referees for very useful comments which helped us better convey our findings.
\end{acknowledgments}

\bibliography{MglineBOAT2}{}
\bibliographystyle{aasjournal}

\end{document}